\documentstyle{article}
\begin{document}
\let\la=\label
\def\nn{\nonumber}
\def\ni{\noindent}
\let\bm=\bibitem
\def\bd{\begin{document}}
\def\ed{\end{document}}
\def\be{\begin{equation}}
\def\ee{\end{equation}}
\def\ba{\begin{array}}
\def\ea{\end{array}}
\def\bea{\begin{eqnarray}}
\def\eea{\end{eqnarray}}
\newcommand{\eq}[1]{(\ref{#1})}

\def\ul{\underline} \def\un{\underline} 
\def\-{{\hskip 1.5pt}\hbox{-}}
\def\fracmm#1#2{{{#1}\over{#2}}} 

\def\low#1{{\raise -3pt\hbox{${\hskip 1.0pt}\!_{#1}$}}}
\def\STr{\, \hbox{STr}\,}
\def\alephnull{~$\large{\aleph_0}\,$~}
\def\sdet{\,\hbox{sdet}\,}
\def\hata{\hat a} \def\hatb{\hat b} \def\hatc{\hat c} 
\def\hatd{\hat d} \def\hate{\hat e} \def\hatf{\hat f} 

\def\pl#1#2#3{Phys.~Lett.~{\bf B#1}(19{#2}){#3}}
\def\np#1#2#3{Nucl.~Phys.{\bf #1B}(19{#2}){#3}}
\def\pr#1#2#3{Phys.~Rev.~ {\bf D#1}(19{#2}){#3}} 
\def\ap#1#2#3{Ann.~of~Phys.~{\bf #1}(19{#2}){#3}}
\def\jmp#1#2#3{Jour.~Math.~Phys.~{\bf #1}(19{#2}){#3}}

\title{\bf Chern-Simons Supersymmetric Branes}  
\author{{\it Pablo Mora}\\
{\it Department of Physics} \\  
{\it University of Maryland at College Park} \\  
{\it College Park, MD 20742-4111, USA} \\  
{\it and}\\
{\it Instituto de Fisica, Facultad de Ciencias}\\ 
{\it Igu\'a 4225, Montevideo, Uruguay}\\}  

\maketitle

\begin{abstract}

The purpose of this Letter is to continue 
the study of the class of models
proposed in the previous paper \cite{mn} hep-th/0002077.
The model corresponds to a system of branes of
diverse dimensionalities with Chern-Simons
actions for a supergroup, embedded in a background 
described also by a Chern-Simons action. The model treats 
the background and the branes on an equal footing,
providing a 'brane-target space democracy'. 
Here we suggest some possible extensions of the 
original model, and disscuss its equations of motion,
as well as the issue of currents and charges carried by the branes.
We also disscuss the relationship with M-theory and Superstring theory. 

\end{abstract}

\centerline{\bf 1.~~Introduction} 

Chern-Simons Supergravity (CSS) theories in diverse dimensions have 
been studied in many papers during the last years 
\cite{witten1,chamseddine, banados1,troncoso, zanelli, banados2,horava}.
Those theories are very interesting on their own right, having a wealth of
nice properties, ranging from being topological (in the sense of being
independent of any background metric), being true gauge theories for extensions
of the standard space-time symmetry groups (like the Poincar\'e 
or anti de Sitter group), having coupling constants that are not 
renormalized
(so that the classical action is also the quantum effective action) 
and being
exactly solvable in 2+1 dimensions. Some time ago 
Chamseddine \cite{chamseddine}
suggested that this kind of models might be regarded as the basis for
an approach to the unification of the fundamental interactions alternative
to the Superstring Theory \cite{superstring} program. More recently
Horava \cite{horava} proposed that a CSS may correspond to the M-theory
\cite{witten2,hull1,townsend1,townsend2},
the fundamental theory underlying the five known consistent 
superstring theories, which would then be an ordinary field theory
(against the belief of most experts on the field).

On other line of development it was attempted  
to recast the 1+1 dimensional 
superstring actions as 2+1 CS theories \cite{mile,kogan} (see also
\cite{witten1}) 
by a sort of 'thickening of
the world sheet',
as a way to benefit from the good properties of the later.

Recently we suggested \cite{mn} a way to introduce 
fundamental supersymmetric
extended objects on CSS, inspired on the model of ref.\cite{dixon}
for the coupling of branes to Yang-Mills fields, 
which brought together these approachs.  
There is a model of that class which has at least two of the 
superstring theories (IIA and IIB) as sectors
of its phase space and it describe branes with actions of the same 
form that the one describing the background, which is a CSS with the 
branes acting as sources for the (super)gauge fields, and interacting 
with that background. Also it is plausible that standard supergravity 
approximately describes some regime of the theory 
\cite{troncoso,horava,banados2}. It was then argued in \cite{mn}
that a model of the kind considered there might correspond to the
quantum effective action of the M-theory.   

In this paper I continue the study of that kind of models by
reviewing its invariances and transformation properties,
discussing its equations of motion and the issue of currents
and charges carried by the branes. Finally I disscuss several
of the questions and open problems where I believe further developments
are to be expected. The model of this paper differs from the one
on ref.\cite{mn} in the fact that we point out that 
it may not be necessary
to introduce a kinetic term 'by hand' 
in order to make contact with superstring
theory.\\ 

\centerline{\bf 2.~~The Action}


The mathematical tools used through this paper are discussed in
(and were to a great extent developed in) 
references.\cite{stora,zumino,manes,alvarez}. We will follow the notation
and conventions of ref.\cite{alvarez}\footnote{An excellent recent work covering the mathematical background
of these articles is the book on anomalies by 
R. Bertlmann \cite{bertlmann}. The reader may find amusing
to know that Dr. Bertlmann socks have inspired in J.S. Bell 
some deep reflections
on the foundations of quantum mechanics \cite{bell}. Bertlmann 
reminiscences of
J.S. Bell on the topic of anomalies (where he was one of the pioneers)
can be found in the Preface to \cite{bertlmann}.}.
We consider a (super)group $~G$~ with generators $~T^I$, 
the gauge potential 1-form $~A=A_m^I T^I \, dx^m $~ and 
the curvature 2-form 
$~F=dA+A^2$, defined on a manifold $~M^{2N+2 }$~ of even dimension $2N$. 
The covariant derivative is $D=d+[A,]$ and it follows that
$F$ satisfy the Bianchi identity
$DF=0$. 
An invariant polynomial $~P(F)$~ is defined as the formal sum
\begin{equation}
P(F)=\sum_{n=0}^{N} \alpha _n \STr \big( F^{n+1} \big) ~~,   
\end{equation} 
where $~\STr \big( T^{I_1}\dots T^{I_{n+1}} \big) 
= g^{I_1\cdots I_{n+1}}$~ stands for an
invariant  symmetric supertrace on the algebra  of $~G$.
An important property is 
\be
d\STr (~~~)=\STr (D~~~)
\ee
We define the action
for our system by  
\be
S=\sum_{n=0}^{N}\alpha _n \int_{S^{2n+1}}k_{01}\STr \left( F^{n+1} \right) 
\ee
where $S^{2N+1}$ is the boundary of $M^{2N+2}$, 
$S^{2N+1}\equiv\partial M^{2N+2}$, and the manifolds $S^{2n+1}$
are embedded into $S^{2N+1}$. In case $S^{2N+1}$ itself
has a boundary $\Omega ^{2N}$, then $S^{2N+1}$ is included 
into the boundary of 
$M^{2N+2}$. 
The manifolds $S^{2n+1}$ may have boundaries 
in manifolds $\Omega ^{2n}$.


The Cartan homotopy operator acting on polynomials 
${\cal P}(F_t,A_t)$, with $A_t$ interpolating between two gauge
potentials $A_0$ and 
$A_1$ as
\be
A_t=t A_1+(1-t)A_0~~,~~F_t=d A_t+A_t^2  ~~.  
\ee
is defined to be
\be
k_{01}{\cal P}(F_t,A_t) =\int_0^1 dt~l_t{\cal P}(F_t,A_t) ~~,  
\ee
with
the operator~$l_t$~ defined to act on arbitrary polynomials by 
\be
l_t A_t=0~~,~~~~ 
l_t F_t = A_1 - A_0\equiv J ~~, 
\ee
and the convention that ~$l_t$~ 
is defined to act as an antiderivation.
 
The dynamical variables are the gauge potentials $~A_m^I$,
the embedding coordinates of the 
submanifolds $S^{2n+1}$,  
$~X^m_{(2n+1)}(\chi _{(2n+1)}^i)$, $m=0,~...,2N+1$ 
where the $~\chi_{(2n+1)}^i$~ with $i~=~0,~...,~2n+1$~,   
are local coordinates 
in $~S^{2n+1}$ and the embedding coordinates of the 
submanifolds $\Omega ^{2n}$, $~X^m_{(2n)}(\xi _{(2n+1)}^i)$, 
$m=0,~...,2N+1$ 
where the $~\xi_{(2n)}^i$~ with $ i~=~0,~...,~2n$~,   
are local coordinates in $~\Omega^{2n}$ 
(of course in the boundary $\Omega ^{2n}$ of $S^{2n+1}$ the
$X^m$'s of the same point must coincide as functions of the 
$\chi$'s or the corresponding $\xi$'s).  
Notice that all these manifolds are supposed to be non compact
at least  in what would be the `time direction'.  It was noticed in
Refs.\cite{zanelli} that the dimensionless coefficients 
$~\alpha_n$~ can
consistently take only a  
discrete set of values if we require
that the quantum theory must  be independent of the way in which the
manifolds $~S ^{2n+1}$~ could be extended into manifolds 
$~M^{2n+2}$~ included into $~M^{2N+2}$~ such that 
$S^{2n+1}\equiv\partial M^{2n+2}$. 
For instance $P(F)=\STr[e^{i F/(2\pi)}]$ would work.\\
A useful relationship is the Cartan homotopy formula 
\be
\big( k_{01}d+dk_{01}\big) {\cal P}(F_t,A_t)
	={\cal P}(F_1,A_1)-{\cal P}(F_0,A_0) ~~. 
\ee
which follows by integrating 
\be
\big( l_td+dl_t\big) {\cal P}(F_t,A_t)
=\frac{\partial}{\partial t}{\cal P}(F_t,A_t)
\ee
over $~t$~ from $~0$~ to $~1$.\\
Then we get
\be
{\cal L}_{2n+1}\equiv k_{01}\STr \left(F_t^{n+1}\right)=
(n+1)\int_0^1dt~\STr \left(JF^{n}\right)
\ee
and 
\be
S=\sum_{n=0}^N\alpha _n\int_{S^{2n+1}}{\cal L}_{2n+1}\equiv
\sum_{n=0}^N \alpha _n S_{2n+1}
\ee
or
\be
S=\sum_{n=0}^N(n+1)\alpha _n 
\int_{S^{2n+1}}\int_0^1dt~\STr \left(JF^{n}\right)
\ee

In ref.\cite{mn} a kinetic term 
was added by hand in the boundary $\Omega ^{2n}$ of 
the manifolds $S^{2n+1}$ given by
\be
S_K^{(2n)}=\frac1 2 \int_{\Omega ^{2n}} d^{2n} \xi_{_{(2n)}} 
     {\sqrt{-\gamma_{_{(2n)}}}}
     \left[ \gamma_{_{(2n)}}^{ij} \STr \left( J_i J_j \right) 
	- (2n-2)  \right] 
\ee
or alternatively of the Born-Infeld-like form
\be
\int_{\Omega ^{2n}} d^{2n}\xi_{_{(2n)}} 
 \STr \left[\sqrt{-\sdet\big\{ J_iJ_j+(F_0)_{ij}+(F_1)_{ij}\big\} }\right]
\ee
or
\be
\int_{\Omega ^{2n}} d^{2n}\xi_{_{(2n)}} \STr \left[\sqrt{-\sdet\big\{ \STr (J_i
J_j) +(F_0)_{ij}+(F_1)_{ij}\big\} }\right] 
\ee
where the superdeterminant is taken in the curved indices $~i~,~j~$
of the  pull-backs on ~$S^d$~ while the supertraces are taken on the 
group indices. We will ignore those kinetic terms here, even though it may
be that they appear as quantum corrections to the action of eq.(3) in the 
quantum effective action.


From the Cartan homotopy formula for 
${\cal P}(F_t,A_t)=I^0_{2n+1}(F_t,A_t)$, 
where the Chern-Simons (CS) form $I^0_{2n+1}(F,A) $ is defined as
\be
I^0_{2n+1}(F,A) =(n+1)\int_0^1 d s~\STr \left(AF_s^{n}\right) ~~,  
\ee
where
\be
A_s=sA~~,~~F_s=d A_s+A_s^2=sF+s(s-1)A^2  ~~.  
\ee
we get 
\be
k_{01}\STr (F_t^{n+1})=I^0_{2n+1}(F_1,A_1) -I^0_{2n+1}(F_0,A_0)
-d\left[k_{01}I^0_{2n+1}(F_t,A_t)\right]
\ee
Where we used that $dI^0_{2n+1}(F,A)=\STr (F^{n+1})$.
The last term is a boundary term which is given explicitly by
\be
C_{2n}(F_1,A_1;A_0,F_0)\equiv -n(n+1)\int_0^1 d s~\int_0^1 d t~s~
\STr \left(A_tJF_{st}^{n-1}\right) 
\ee
with $F_{st}=sF_t+s(s-1)A_t^2$ and $A_t=tA_1+(1-t)A_0$.\\

\centerline{\bf 3.~~Invariances of the Action}

By construction the action of eq.(3) and the kinetic terms
given above are generally covariant.

The content of this section regarding the transformation properties
of CS forms and descent equations was developed in the context
of anomalies in quantum field theories in 
refs.\cite{stora,zumino,manes,alvarez}. 
I find it worthwhile to review
it here because I apply it in a different  context and 
conceptual framework.

Under (super)gauge transformations we have
\be
A_r^g=g^{-1}(A_r+d)g~~,~~r=0,1
\ee
where $g$ is an element of the group. It follows that
$~J=A_1-A_0$~ transforms covariantly if both $A_1$
and $A_0$ are transformed with the same $g$
\be
J^g=g^{-1}Jg 
\ee
Also
\be
F^g=g^{-1}Fg
\ee
Under infinitesimal
gauge transformations 
\be
\delta_{v} A_r=dv +[ A_r,v ]~~,~~r=0,1
\ee
Then
\be 
\delta _{v} J = [ J,v ]~~.
\ee
and
\be 
\delta _{v} F= [ F,v ]~~.
\ee
From the facts that $F$ and $J$ are gauge covariant, 
the ciclicity of $\STr ()$ and eq.(11) it follows that the action 
and also the kinetic terms of eqs.(12-14) are gauge 
invariant.\\
In order to compute the change of the action under  gauge 
transformations involving only one of the gauge fields $A_1$
or $A_0$ it is useful to consider elements of the gauge group
$g(x,\theta )$ function of the point $x$ on the base manifold
and a set of parameters $\theta ^\alpha$ on some 'parameter space',
such that $g(x,\theta =0)=1$ (the identity). In addition to the 
standard exterior derivative 
$d=dx^{\mu}\frac{\partial}{\partial x^{\mu}}$ we define the exterior 
derivative in parameter space 
$\zeta =d\theta ^{\alpha}\frac{\partial}{\partial \theta ^{\alpha}}$.
If
\be
\overline{A}=g^{-1}(A+d)g=A^g
\ee 
and
\be
{\cal A}=g^{-1}(A+d+\zeta )g=g^{-1}(A+\Delta )g=\overline{A}+v 
\ee
with $\Delta =d+\zeta $ and $v=g^{-1}\zeta g$. We have 
$d^2=\zeta ^2=d\zeta +\zeta d=\Delta ^2=0$. It is easy to verify
that $\zeta \overline{A}=-D_{\overline{A}}v$ so that $\zeta$ generates 
gauge transformations with parameter 
$v_{\alpha}=g^{-1}\frac{\partial}{\partial \theta ^{\alpha}}g$.
The derivative $\zeta$ corresponds to the BRS operator 
and $v$ to the Fadeev-Popov
ghost\cite{stora,zumino,manes,bertlmann}. Defining
$\overline{F}=d\overline{A}+\overline{A}^2$ and 
${\cal F}=\Delta {\cal A}+{\cal A}^2$  it is possible to check the
'Russian formula'
\be
{\cal F}({\cal A})=\overline{F} (\overline{A} )=g^{-1}F(A)g
\ee
Considering
$${\cal A}_t=t{\cal A}_1+(1-t)A_0$$
$$\overline{A}_t=t\overline{A}_1+(1-t)A_0$$
and $\zeta A_0=0$, then from the 'Russian formula' and
the Cartan homotopy formula  with ${\cal P}_t=\STr (F_t^{n+1})$
for ${\cal A}_t$ and $\overline{A}_t$ we get
\be
(d+\zeta)I^0_{2n+1}(\overline{A}_1+v,A_0)=
dI^0_{2n+1}(\overline{A}_1,A_0)
\ee
where $I^0_{2n+1}(A_1,A_0)=k_{01}\STr (F_t^{n+1})={\cal L}_{2n+1}$
correspond to the pieces of diverse dimension of our lagrangian. If
we expand by the order in $v$
\be
I^0_{2n+1}(\overline{A}_1+v,A_0)=
\sum_{k=0}^{2n+1}I^k_{2n+1-k}(v,\overline{A}_1,A_0)
\ee
then we obtain the 'descent equations'
\be
\zeta I^{k}_{2n+1-k}(v,\overline{A}_1,A_0)+
dI^{k+1}_{2n-k}(v,\overline{A}_1,A_0)=0~~,~~k=0,~...,2n+1
\ee
In particular
\be
\zeta I^{0}_{2n+1-k}(\overline{A}_1,A_0)+
dI^{1}_{2n-k}(v,\overline{A}_1,A_0)=0
\ee
gives the variation of our action under a gauge transformation
involving only $A_1$ 
(notice that $\overline{A}_1\mid _{\theta =0}=A_1$) as a boundary term.
 A similar 
identity holds for gauge transformations involving only $A_0$.\\

\centerline{\bf 4.~~Equations of Motion }

In the case of CS gauge theory or supergravity 
without branes or boundaries
the action is  
\be
S=\int_{S^{2n+1}}I_{2n+1}^0(F,A)=\int_{M^{2n+2}}\STr (F^{n+1})
\ee
then under variations of the gauge potential
\be
\delta S= (n+1)\int_{M^{2n+2}}\STr (D(\delta A)F^{n}) =
(n+1)\int_{M^{2n+2}}d\left[\STr (\delta AF^{n})\right] 
\ee
where we used $\delta F=D(\delta A)$ and eq.(2). From Stokes theorem
\be
\delta S= (n+1)\int_{S^{2n+1}}\STr (\delta AF^{n})  
\ee
Then the equations of motion $\frac{\partial S}{\partial A}=0$ are
\cite{chamseddine,banados1,troncoso}
\be
\STr (T^IF^{n})=0  
\ee
Whether or not these equations are related to General Relativity and/or 
standard Supergravity in diverse dimensions has been disscussed in 
Ref.\cite{witten1, chamseddine, banados1, troncoso, horava, banados2}.

In the case there are boundaries and branes we need to use that
$$\delta _1 J =\delta A_1~~~,~~~\delta _0 J = -\delta A_0$$
$$\delta _r F_t = D_t ( \delta _r A_t ) =d( \delta _r A_t ) 
+[A_t,( \delta _r A_t ) ]~~~,~~~r=0,1$$
$$\delta _1 A_t= t \delta A_1~~~,~~~\delta _0 A_t = (1-t)\delta A_0$$
Therefore we have for variations of $A_1$
\be
\delta _1 {\cal L}_{2n+1}=(n+1)\int_0^1dt \STr (\delta A_1F_t^{n})  
+n(n+1)\int _0^1dt~t\STr( JD_t(\delta A_1)F_t^{n-1})
\ee
but
\be
d\left[\STr (\delta A_1F_t^{n})\right] =\STr (D_t J\delta A_1F_t^{n-1})  
-\STr (JD_t(\delta A_1)F_t^{n-1})    
\ee 
where we used $d\STr (~~)=\STr (D_t~~)$ and the Bianchi identity $D_tF_t=0$,
then
\begin{eqnarray}
\delta _1 {\cal L}_{2n+1}=(n+1)\int_0^1dt \STr (\delta A_1F_t^{n})  
+n(n+1)\int _0^1dt~t\STr(\delta A_1 D_t(J)F_t^{n-1})\nonumber \\ 
+d\left[n(n+1)\int _0^1dt~t\STr( \delta A_1 J F_t^{n-1})\right]
\end{eqnarray}
 The last term of the second member is a boundary term. Under variations
of $A_0$ we have  
\begin{eqnarray}
\delta _0 {\cal L}_{2n+1}=-(n+1)\int_0^1dt \STr (\delta A_0F_t^{n})  
+n(n+1)\int _0^1dt~(1-t)\STr(\delta A_0 D_t(J)F_t^{n-1})\nonumber \\ 
+d\left[ n(n+1)\int _0^1dt~(1-t)\STr( \delta A_0 J F_t^{n-1}) \right]
\end{eqnarray}
If we write
\be
\delta _r{\cal L}_{2n+1}=\STr (\delta A_r Q^{(r)}_{2n} )
+d\left[\STr (\delta A_r R^{(r)}_{2n-1} )\right]
\ee
where
\begin{eqnarray} 
Q^{(1)}_{2n} =(n+1)\int_0^1dt  F_t^{n}  
+n(n+1)\int _0^1dt~t  D_t(J)F_t^{n-1}\nonumber\\
Q^{(0)}_{2n}=-(n+1)\int_0^1dt  F_t^{n}  
+n(n+1)\int _0^1dt~(1-t)  D_t(J)F_t^{n-1}\nonumber\\
R^{(1)}_{2n-1}=n(n+1)\int _0^1dt~t J F_t^{n-1}\nonumber\\
R^{(0)}_{2n-1}=n(n+1)\int _0^1dt~(1-t) J F_t^{n-1} 
\end{eqnarray}

Then we can write

\be
\delta _r S=\sum_{n=0}^N\alpha _n
\left[ \int_{S^{2n+1}}\STr (\delta A_r Q^{(r)}_{2n} )
+\int_{\Omega ^{2n}}
\STr (\delta A_r R^{(r)}_{2n-1} )\right]
\ee
or
\be
\delta _r S=\int_{S^{2N+1}}d^{{\small 2N+1}}x~ \delta _r~A_m^I ~J^{(r)mI}
\ee
where
\be
J^{(r)mI}(x^m) = \sum_{n=0}^N\alpha _n\bigg[
\int_{S^{2n+1}}d^{2n+1}\chi _{2n+1} {\cal J}_{(2n+1)}^{(r)mI} 
+\int_{\Omega ^{2n}}d^{2n}\xi _{2n}
{\cal J}_{(2n)}^{(r)mI} \bigg]
\ee
with

\begin{eqnarray}
{\cal J}_{(2n+1)}^{(r)mI}= 
\delta ^{2N+1}(X_{(2n+1)}(\chi _{2n+1})-x^m)
\STr \left( T^I (Q^{(r)}_{2n}) _{m_2...m_{2n+1}} \right)
\times\nonumber\\
\times\partial _{i_1}X^{[m}_{(2n+1)} 
\partial _{i_2}X^{m_2}_{(2n+1)}... 
\partial _{i_{2n+1}}X^{m_{2n+1}]}_{(2n+1)}
\epsilon ^{i_1...i_{2n+1}}
\end{eqnarray}
and
\begin{eqnarray}
{\cal J}_{(2n)}^{(r)mI}= 
\delta ^{2N+1}(X^m_{(2n+1)}(\xi _{2n})-x^m)
\STr \left( T^I (R^{(r)}_{2n-1}) _{m_2...m_{2n}} \right)
\times\nonumber\\ 
\times\partial _{i_1}X^{[m}_{(2n)} 
\partial _{i_2}X^{m_2}_{(2n)}... 
\partial _{i_{2n}}X^{m_{2n}]}_{(2n)}\epsilon ^{i_1...i_{2n}} 
\end{eqnarray}
The equations of motion $\frac{\partial S}{\partial A_r}=0$  
are then
\be
J^{(r)mI} = 0
\ee
These equations can be interpreted as the equations found
before in the case there are no boundaries or branes but now
with source terms given by currents carried by the branes.

Concerning the equations of motion corresponding to extremize the
action under variations of the embedding functions $X$ it is 
convenient to write
\begin{eqnarray}
S=\sum_{n=0}^N\alpha _n
\bigg[ \int_{S^{2n+1}} d^{2n+1}\chi _{2n+1} ~(\omega _{2n+1})_{m_1...m_{2n+1}} 
\partial _{i_1}X^{[m_1}_{(2n+1)} 
... 
\partial _{i_{2n+1}}X^{m_{2n+1}]}_{(2n+1)}\epsilon ^{i_1...i_{2n+1}}  
\nonumber\\
+\int_{\Omega ^{2n}} d^{2n}\xi _{2n}~
(\omega _{2n})_{m_1...m_{2n}}\partial _{i_1}X^{[m_1}_{(2n)} 
... 
\partial _{i_{2n}}X^{m_{2n}]}_{(2n)}\epsilon ^{i_1...i_{2n}}  \bigg]
\end{eqnarray}
where we separated the bulk and boundary 
contributions to ${\cal L}_{2n+1}$. In the previous expresion
the dependence of $S$ on the functions $X$ is through the $\omega$'s
while the dependence of $S$ on $\partial X$ is through the pull-back
factors. The Euler-Lagrange equations for $X_{(p)}^s$then give
\be
\bigg[p\frac{\partial}{\partial X^r_{(p)}}(\omega _{(p)})_{ sm_2...m_p}
-\frac{\partial}{\partial X^s_{(p)}}(\omega _{(p)})_{r m_2...m_p}\bigg]
\partial _{i_1}X^{[r}_{(p)} 
\partial _{i_2}X^{m_2}_{(p)} 
... 
\partial _{i_{p}}X^{m_{p}]}_{(p)}\epsilon ^{i_1...i_{p}}=0  
\ee   
In applying the Euler-Lagrange for the 'bulk' $S^{2n+1}$ we left out
a boundary term
\be
\partial _{i_1}\bigg[(\omega _{(p)})_{ sm_2...m_p}
\partial _{i_2}X^{[m_2}_{(p)} 
... 
\partial _{i_{p}}X^{m_{p}}_{(p)}\epsilon ^{i_1...i_{p}}
\delta X^{s]}_{(p)}\bigg]
\ee   
We can require the boundary term term to vanish,
in analogy with open strings,
\be
(\omega _{(p)})_{ sm_2...m_p}
\partial _{i_2}X^{[m_2}_{(p)} 
... 
\partial _{i_{p}}X^{m_{p}}_{(p)}\epsilon ^{i_1...i_{p}}
\delta X^{s]}_{(p)}=0
\ee   
which is not to be taken as a condition on what points  
can be swept by the boundaries of the branes, its velocities
or the allowed variations $\delta X$ but only on the spatial derivatives
of the functions $X$ at the boundary. Alternatively we can add that term
as an extra contribution to the Euler-Lagrange equations at the boundary.

If we add the kinetic terms of eq.(12-14) there would be 
an extra term to the
current located on the boundaries $\Omega ^{2n}$ 
of the branes, and extra 
terms in the Euler-

Lagrange equations. The equations of motion of
the auxiliary metrics $\gamma$ in the kinetic 
terms of eq.(12) are algebraic.\\

\centerline{\bf 5.~~Discussion}

{\it Connection with Superstring/M-theory}

In ref.\cite{mn} we considered a the model of eq.(3) plus
the kinetic term of eq.(12) added 'by hand' for the
M-theory group $~OSp(32|1)$~\cite{holten,townsend2,horava} 
and related it to IIA and IIB
superstrings. 
That group has generators $~P_a$ (translations), $~Q_{\alpha}$
(generators of supersymmetries), 
$~M_{ab}$~(Lorentz) and $~Z_{a_1...a_5}$, $a=0,...10$.
We took the symmetric trace
to be the standard symmetricized supertrace in the adjoint
representation of $~G$\footnote{For the following discussion the relevant
traces of products of generators in the adjoint representation of
$~OSp(32|1)$~ are \cite{mn} $\STr (P P) $, $\STr (P Q)$~ and $\STr (Q Q )$,  while
for the WZNW-term the relevant traces are $\STr (P P P) $, $\STr (P P Q)
$, $\STr (Q Q P) $~ and  $\STr (Q Q Q)$.  Among these, the non-vanishing
ones (after the limiting process of \cite{mn})are normalized as 
$$ \STr (P_a P_b) = \eta _{a b}~~,~~~~
	\STr (P_a Q _{\alpha}Q_{\beta} )
	=  \left(\gamma _a~C^{-1}\right) _{\alpha\beta} $$}.

The point we would like to make here is that 
that connection can be made for the action of eq.(3) alone.
We consider as our candidate to the M-theory action
\be
S=\sum_{n=1}^{5}\int_{S^{2n+1}}k_{01}
\STr \left( exp\left(i\frac{F}{2\pi}\right) \right) 
\ee
for $~OSp(32|1)$. The integrals in the previous expression are
supposed to pick up the differential forms of the right order.
We will use the same notation as \cite{mn} and take the 
Inonu-Wigner limit as it was done in that paper.
We need to consider a pure gauge $A_1=g^{-1}dg$ 
with $g$ restricted
to the form
$g=e^{i\hat{X}^aP_a+\theta ^{\alpha}Q_{\alpha}}$.
Then esentially $A_0=[dX^a+i\overline{\theta}\gamma ^ad\theta]
+d\theta ^{\alpha}Q_{\alpha} $. We will also make $A_0=-A_1$.
Then proceeding as in \cite{mn} we consider a 11D slab with two
10D noncompact boundaries (each with the topology of $R^{10}$) 
for which 
we identify the gauge parameters 
$\hat{X}^a$ with the 10D
coordinates. We will also take  the  pull-back of the
gauge superfield $A_1$ to be 
self-dual or anti-self dual $A_1=\pm *A_1$
with respect to some
arbitrary auxiliary metric in the 2D boundaries 
of the 2-brane, which are
contained in the 10D boundary of the 11D slab.
To fix ideas we may think that metric 
is a 2D Minkowski metric
$\eta =(-1,1)$, then the (anti)self-duality 
condition means that we are keeping
only the (left)right movers with respect to that metric.
The bulk parts Chern-Simons of eq.(52) for the 
2-brane give boundary
WZW terms because the potentials are pure 
gauge, and both terms actually 
add because we chose $A_1=-A_0$. On the other 
hand from eq.(18)
$C_2=\STr (A_1A_0)$, but the wedge product of a 
differential form with
its dual with respect to a metric is the square 
with that metric. It follows
that the $C_2$ in each side of the slab 
would look like half
the kinetic term 
computed with that metric (in the sense we 
would only have right or
left movers). The 11D Majorana spinor has 32
components. We can split those 32 components in 
10D either as    
$~({\bf 32})
= ({\bf 16}_L, {\bf 16}_R)$ or as $~({\bf 32})
= ({\bf 16}_R, {\bf 16}_R)$ (equivalently $~({\bf 32})
= ({\bf 16}_L, {\bf 16}_L)$ ), where $~{_L}$~ and $~{_R}$~ 
denote the 
chiralities  in 10D.     
Chosing properly the anti-self duality or self duality 
conditions so that we have for instance left movers in one face 
and right movers in the other we can assemble 
left or right movers for $X$ and for the spinors from both faces.
Then each choice of the splitting of 11D spinors into two 10D
spinors yield IIA or IIB superstrings. 
Of course the previous considerations only means that IIA
and IIB strings corresponds to 'sectors 
of the phase space' ofthe theory,
contibuting to the quantum path integral. 
It would take more work to 
check if those configurations are 
actually solutions of the equations
of motion. 

{\it Duality} 

We can distinguish between the base
manifold coordinates $X$ and the gauge parameters $\hat{X}$ ,
and only identify them as a coordinate choice after choosing
a topology for the base manifold, as done in a particular
case in \cite{mn}. That means that we can get the various dualities
of \cite{bergshoeff} at the level of the $OSp(1,32)$ algebra,
corresponding to different choices of the set of operators
associated to translations in 10D, by picking different $g$'s
on the Maurer-Cartan form as above with the proper $P_a$, and
doing the identification $X\equiv\hat{X}$ for the corresponding 
$\hat{X}$ on the appropiate 10D submanifold.

Concerning T-duality, the fact that the dimensional reduction of 
Chern characters are Chern characters in the lower dimension would
single them out of all the possible invariant polynomials. Namely
we should consider just a linear combination of terms of the form
$\STr (F^{n+1})$ instead of
some arbitrary
combination of products of traces or 
some supersymmetric extension of the Euler
characteristic. 
 
{\it D-branes and K-theory}

Several recent works deal with the issue of D-branes and K-theory
\cite{minasian,witten3}. The situation is often stated as 'K-theory
is to be preferred to cohomology' or equivalently 'gauge fields are to
be preferred to $p$-form RR-fields'. In our model, as pointed out
in \cite{mn}, we can mimic the RR-fields (which would then be 
regarded as composite) with the CS forms of one of 
the gauge potentials (say $A_1$) which would couple 
(with an 'anomalous coupling') to the other one (say $A_0$). The
'anomalous gauge transformation rule' of the RR field is then built in.
Also K-theoretic constructions 
involve a doubling of the gauge fields, as our model does. 
It seems therefore reasonable to investigate the 
relationship between both
approachs.

{\it Quantum theory and quantum effective action}

The quantum theory is formaly defined by the path integral
$$Z=\sum_{topologies}\sum_p\int{\cal D}A~{\cal D}X_{(p)}~e^{iS/\hbar}$$ 
where it is undertood that we must sum over all the gauge field 
configurations and brane and base space geometries and topologies.
Suitable gauge fixind procedures should be used to eliminate redundancies
in summing configurations corresponding to the same physical state.
The topological character of the action and the quantization of the 
coupling constants may be taken to imply that the action of eq.(3)
is already the quantum effective action. However a careful analysis
of this question is clearly required.

{\it Anomalies and full gauge invariance}

It is worthwhile to notice that the variation of our action under
a gauge transformation involving only
one of the gauge fields has the right form to be canceled by
an anomaly on the boundary $\Omega ^{2n}$ \cite{manes}, as it satisfies
the Wess-Zumino consistency condition and comes from a chain 
of 'descent equations'. It is therefore tempting to look for such a
mechanism to ensure the full gauge invariance 
of the quantum theory
under arbitrary gauge tranformations of 
$A_0$ or $A_1$\footnote{That seems
also desirable as making the same variation on both sides of a 
't-parameter space slab' is reminiscent of 'distant parallelism',
even though actually both $A_0$ and $A_1$ are evaluated at the same 
space-time point. Incidentally I wonder if this additional one dimensional
t-parameter space has anything to do with F-theory \cite{vafa}.}. However a priori it does not seem that we need to
have anomalies at all on the even dimensional brane boundaries, as we
could in principle arrange the fermion fields (the fermionic gauge 
fields of the supergroup) in couples of opposite chirality so that 
the world volume theory is non-chiral. 
Yet if we chose to make it chiral,
computing the precise form of the 
anomaly would be highly non-trivial,
as we have {\it fermionic gauge fields} 
transforming as such under gauge 
transformations and reparametrizations 
(general coordinate transformations).
It follows that the usual formulas for 
'standard' (wether gauge or 
gravitational) or 'sigma model' 
anomalies would not apply. 
A discussion
of anomalies for extended objects 
(which unfortunately I could not  translate in 
any straightforward way to the problem at hand),
and when and how to apply which formulas
can be found in ref.\cite{izquierdo} 
and \cite{duff}\footnote{See also
\cite{cheung} and references therein. 
There are more recent references on 
brane anomalies but do not apply to 
the peculiarities of our model.}.
This approach seems to be the best hope to single 
out the gauge supergroup and
the dimension of the space-time 
manifold\footnote{It would be ironic if a model of the
kind studied in \cite{mn} and the present paper is 
relevant to M-theory as I believe it is, 
as the action of the model itself
is essentially an 'anomaly'.}. 
If our model in eleven dimensions and
with gauge group $~OSp(32|1)$ has as limiting cases 
the the five consistents
superstring theories, then consistency  and anomaly cancelation 
considerations from the later should translate into the fact that 
the former is the only one of our class of models that works. We can
make a 'hand waving' argument in the sense that a fully consistent 
theory of Nature should be 'perturbatively smooth' when expanded 
around any 'point' of its phase space, in the vague sense that 
each order must be finite, even if the 'point' is not the true 
vacuum and the whole series does not converge. 
  
{\it Vacuum and Phenomenology}

If the action is alredy the effective quantum action, as claimed,
the problem of finding the vacuum reduces to finding a solution of
the classical equations of motion. Doing phenomenology would require
a realistic solution in the sense of having four large nearly flat
3+1 space-time dimensions (at least at some stage of cosmic evolution)
and the masses and coupling constants of particle physics could be read
from the coefficients of the lower order terms in a background field 
expansion quantization.

{\it Group manifold/Superspace formulation}

An interesting possibility would be to treat the base manifold 
and the fiber on the same footing by a group manifold approach.
That may furthermore 
allow to treat the BRS operator $\zeta$ and the exterior
derivative $d$ on a symmetric fashion, giving rise to a sort of 'double
group manifold approach'. It may also be 
possible and useful to extend
the definition of $l_t$ and the one-dimensional t-parameter to 
a manifold
as in \cite{manes} and treat also $l_t$ in a more symmetrical
fashion (a 'triple group manifold approach'?).

{\it Pregeometric theory}

I find very attractive the purely algebraic way in which the
differential structure is treated in Refs.\cite{stora,zumino,
manes,alvarez}. It is also remarkable the contrast between
the simplicity and terseness of the formalism set forth on those
papers and the power and scope of those methods. I believe 
that is a broad hint of the kind of conceptual and mathematical 
framework required to describe a fundamental theory of Nature 
for which the differential structure is a dynamical entity.
It seems that the more fundamental formulation of such a theory
must be a discrete one ('Covariant Matrix Theory')
\cite{banks,moore,west,hull2,gebert}. One can think that
it is not possible to give physical content to the de Rham complex
without giving up the smooth manifold picture of the space-time.   
As pointed in \cite{hull2} the pregeometric approach and the 
geometric approach mentioned in the previous item might not be 
compatible, and we believe the pregeometric one is more likely
to give a conceptually tight picture of physical reality. 
Possibly asking for a pregeometric {\it and} a geometric
group manifold/superspace 
formulation at once may be like asking for Newton's Laws 
{\it and} circular planetary orbits.\\

I am grateful to S.J. Gates Jr. and H. Nishino for 
stimulating discussions as well as support and encouragement.

\end{document}